\documentclass[12pt,a4paper]{article}
\usepackage{anziamjedraft}
\usepackage{amsmath}
\usepackage{amssymb}
\usepackage{wasysym}
\usepackage{verbatim}

\usepackage{color}
\usepackage[english]{babel}
\usepackage{csquotes}

\usepackage{pgfplots}
\usetikzlibrary{arrows,patterns,shapes,positioning,calc}

\usepackage{color}
\usepackage{lipsum}
\usepackage{hyperref}
\hypersetup{
    bookmarks=true,         % show bookmarks bar?
    unicode=false,          % non-Latin characters in Acrobat’s bookmarks
    pdftoolbar=true,        % show Acrobat’s toolbar?
    pdfmenubar=true,        % show Acrobat’s menu?
    pdffitwindow=false,     % window fit to page when opened
    pdfstartview={FitH},    % fits the width of the page to the window
    pdftitle={Polarizablity of 2D and 3D conducting objects using method of moments},    % title
    pdfauthor={Shahpari},     % author
    pdfsubject={polarizability},   % subject of the document
    pdfcreator={Shahpari},   % creator of the document
    pdfproducer={Shahpari}, % producer of the document
    pdfkeywords={polarizability} {Method of Moments}, % list of keywords
    pdfnewwindow=true,      % links in new window
    colorlinks=true,       % false: boxed links; true: colored links
    linkcolor=blue,          % color of internal links
    citecolor=red,        % color of links to bibliography
    filecolor=magenta,      % color of file links
    urlcolor=cyan           % color of external links
}
\title{Polarizability of 2D and 3D conducting objects using method of moments}

\author{Morteza Shahpari}
\address{Centre for Wireless Monitoring and Applications,
School of Engineering, Griffith University, 
Queensland 4111, \textsc{Australia}.}
\mailto{m.shahpari@griffith.edu.au}

\author{D.~V. Thiel}
\address{Centre for Wireless Monitoring and Applications,
School of Engineering, Griffith University, 
Queensland 4111, \textsc{Australia}.}
\mailto{d.thiel@griffith.edu.au}

\author{Andrew Lewis}
\address{Institute for Integrated and Intelligent Systems, Griffith University,
Queensland 4111, \textsc{Australia}.}
\mailto{a.lewis@griffith.edu.au}
\date{(Received 15 November 2012; revised 20 July 2013)}

\begin{document}

\maketitle

% Use the \verb|abstract| environment.
\begin{abstract}
Fundamental antenna limits of the gain-bandwidth product are derived from polarizability calculations. 
This electrostatic technique has significant value in many antenna evaluations.
Polarizability is not available in closed form for most antenna shapes and no commercial electromagnetic packages have this facility. 
Numerical computation of the polarizability for arbitrary conducting bodies was undertaken using an unstructured triangular mesh over the surface of 2D and 3D objects.
Numerical results compare favourably with analytical solutions and can be implemented efficiently for large structures of arbitrary shape.
\end{abstract}

% By default we include a table of contents in each paper.
\tableofcontents

% Use \verb|\section|, \verb|\subsection|, \verb|\subsubsection| and 
% possibly \verb|\paragraph| to structure your document.  Make sure 
% you \verb|\label| them for cross-referencing with \verb|\ref|\,.
\section{Introduction}
\label{sec_intro}

Polarizability is an important parameter in a variety of physical science disciplines including scattering and, molecular and chemical physics.
Recently engineers used this parameter for antenna modelling.
Gustafsson et~al.~\cite{Gustafsson2009_TAP} demonstrated the relationship between the maximum possible antenna gain-bandwidth product and the polarizability of the antenna obstacle.
This is now recognised as a method of calculating a fundamental limit for antennas.
The maximum electromagnetic scattering of a plane wave incident on an obstacle (e.g., metamaterial~\cite{Sohl_2007_JPhysA,Sohl_2007_JAP}, periodic arrays~\cite{Gustafsson2009_EPL}) is related to the static polarizability of the obstacle.
Popular commercial packages for antenna modelling, for example, \textsc{Ansys} \textsc{Hfss}, \textsc{Feko}, \textsc{Awr}, \textsc{Ie3d}, etc do not calculate the polarizability.

In this article we use the method of moments (\textsc{MoM}) to calculate the polarizability of arbitrary geometries.
The \textsc{MoM} technique is commonly used in modelling wire structures for both radiation and scattering problems, assuming a thin wire approximation~\cite{NEC,Harrington_b_Field_Comp}.
In the \textsc{MoM} polarizability calculation, three dimensional structures are modelled using triangular mesh elements.
The technique was implemented in \textsc{Matlab} to calculate the polarizability of arbitrary shaped objects with infinite conductivity (i.e. perfect electric conductors, \textsc{Pec}).
\textsc{Feko}, which is an antenna simulation package, was used to create triangular mesh elements for arbitrary objects and the 3D mesh was exported to the \textsc{MoM} routine in \textsc{Matlab} using the \textsc{Stl} format\footnote{Standard tessellation language \textsc{Stl} is the industrial standard for handling triangulated meshes \url{http://wiki.netfabb.com/STL\_Files\_and\_Triangle\_Meshes}, \url{http://en.wikipedia.org/wiki/STL\_\%28file\_format\%29}
 }.
Triangular mesh elements have the advantage of constructing arbitrary geometries without staircase approximations.

A brief description of the \textsc{MoM} solution is described in Section~\ref{sec_Formulation}.
Section~\ref{sec_Implementation} explains the implementation of the algorithm. Section~\ref{sec_Validation} demonstrates the validity of the code by comparing the \textsc{MoM} results with closed form solutions of simple geometric objects.

\section{Formulation}
\label{sec_Formulation}

In electromagnetics, Laplace's equation is used to describe the electrostatic potential in a charge free region. 
Assuming that the antenna has no  static charge, then Laplace's equation in the integral form is~\cite{Harrington_b_Field_Comp}
\begin{align}
x_{j}+C_{j} = \iint_{S} \frac{\rho_j(\vec{x}')}{4\pi \left | \vec{x}-\vec{x}' \right |} d{S'},
\label{eq_laplace}
\end{align}
where $\rho_{j}$ and $x_j$ are the surface charge density and distance from the origin along $\mathbf{x_j}$ axis when the object is located in a static field of the unit amplitude in the $\hat{x}_{j}$ direction.
This is integrated over the surface of the geometry $S$.
Also, $\vec{x}$ and $\vec{x}'$ refer to observation and source points, respectively.
If the object is asymmetrical or is offset from the origin, the sum of the total induced charge tends to be nonzero. 
This is contrary to the charge conservation law. 
The constant $C_{j}$ is added to ensure the total charge on the object is zero: 
\begin{align}
\iint_{S} \rho_{j}(x)dS=0.
\end{align}

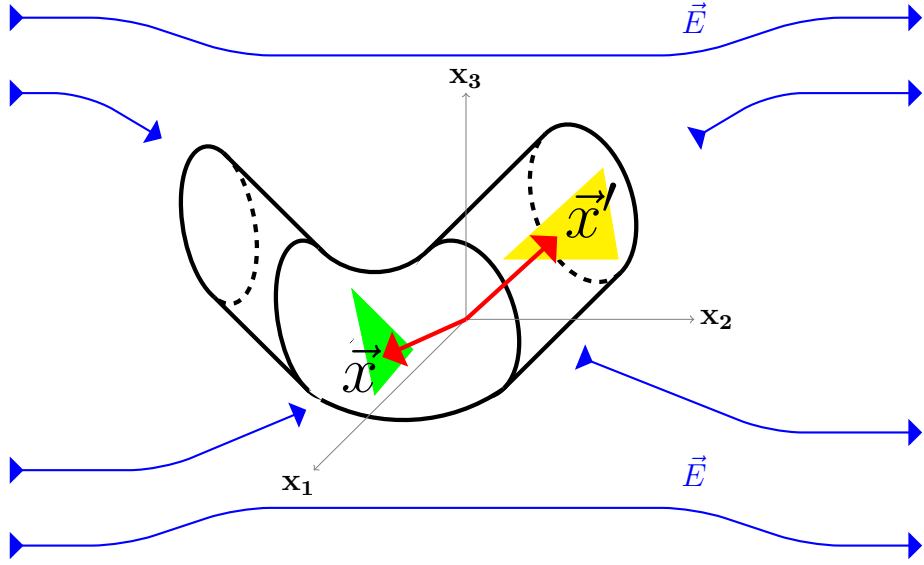
\begin{figure}
\newcommand{\colorgeometry}{white}
\centering
\begin{tikzpicture}

\begin{scope}[x={(.7cm,-.3cm)}]
\pgfmathsetmacro{\length}{4}
\path (1,0,0);
\pgfgetlastxy{\cylxx}{\cylxy}
\path (0,1,0);
\pgfgetlastxy{\cylyx}{\cylyy}
\path (0,0,1);
\pgfgetlastxy{\cylzx}{\cylzy}
\pgfmathsetmacro{\cylt}{(\cylzy * \cylyx - \cylzx * \cylyy)/ (\cylzy * \cylxx - \cylzx * \cylxy)}
\pgfmathsetmacro{\ang}{atan(\cylt)}
\pgfmathsetmacro{\ct}{1/sqrt(1 + (\cylt)^2)}
\pgfmathsetmacro{\st}{\cylt * \ct}
\fill[\colorgeometry] (\ct,\st,0) -- ++(0,0,-\length) arc[start angle=\ang,delta angle=180,radius=1] -- ++(0,0,\length) arc[start angle=\ang+180,delta angle=-180,radius=1];
\coordinate (A1) at (\ct,\st,0);
\coordinate (B1) at (-\ct,-\st,0);

\begin{scope}[every path/.style={ultra thick}]
\draw [fill= \colorgeometry] (0,0,0) circle[radius=1];
\draw (\ct,\st,0) -- ++(0,0,-\length);
\draw (-\ct,-\st,0) -- ++(0,0,-\length);
\draw (\ct,\st,-\length) arc[start angle=\ang,delta angle=180,radius=1];
\draw[dashed] (\ct,\st,-\length) arc[start angle=\ang,delta angle=-180,radius=1];
\end{scope}
\end{scope}

\begin{scope}[xshift=-2cm]

\begin{scope}[x={(.5cm,-.3cm)},z={(.5cm,-.5cm)}]
\pgfmathsetmacro{\length}{2.5}
\path (1,0,0);
\pgfgetlastxy{\cylxx}{\cylxy}
\path (0,1,0);
\pgfgetlastxy{\cylyx}{\cylyy}
\path (0,0,1);
\pgfgetlastxy{\cylzx}{\cylzy}
\pgfmathsetmacro{\cylt}{(\cylzy * \cylyx - \cylzx * \cylyy)/ (\cylzy * \cylxx - \cylzx * \cylxy)}
\pgfmathsetmacro{\ang}{atan(\cylt)}
\pgfmathsetmacro{\ct}{1/sqrt(1 + (\cylt)^2)}
\pgfmathsetmacro{\st}{\cylt * \ct}
\fill[\colorgeometry] (\ct,\st,0) -- ++(0,0,-\length) arc[start angle=\ang,delta angle=180,radius=1] -- ++(0,0,\length) arc[start angle=\ang+180,delta angle=-180,radius=1];
\coordinate (A2) at (\ct,\st,0);
\coordinate (B2) at (-\ct,-\st,0);

\begin{scope}[every path/.style={ultra thick}]
\draw [fill=\colorgeometry] (0,0,0) circle[radius=1];
\draw (\ct,\st,0) -- ++(0,0,-\length);
\draw (-\ct,-\st,0) -- ++(0,0,-\length);
\draw (\ct,\st,-\length) arc[start angle=\ang,delta angle=180,radius=1];
\draw[dashed] (\ct,\st,-\length) arc[start angle=\ang,delta angle=-180,radius=1];
\end{scope}
\end{scope}

\end{scope}

\coordinate (BendCenter) at (-4,-2);
pgfpathmoveto{A1};

\fill [\colorgeometry] (B1) arc[start angle =-45, delta angle =-92, radius=1] -- (A2) -- (B2) -- (B1) -- (A1) arc[start angle =-55, delta angle =-63, radius=2.3] -- (B2) ;

\draw [ultra thick] (A1) arc[start angle =-55, delta angle =-63, radius=2.3]; 

\draw [ultra thick] (B1) arc[start angle =-45, delta angle =-92, radius=1];

%coordinate system
\draw [help lines,->] (0,0) -- (-2,-2);
\node at (-2.2,-2.2) {$\mathbf{x_1}$};
\draw [help lines,->] (0,0) -- (3,0);
\node at (3.3,0) {$\mathbf{x_2}$};
\draw [help lines,->] (0,0) -- (0,3);
\node at (0,3.2) {$\mathbf{x_3}$};
%arbitrary shape
%\draw (-1,2)-- (-0.5,2.3) -- (0.5,2.2)--(1.2,2.7)-- (1.9,2)-- (1.9,0.5)-- (0.7,0.2)-- (0,0.5) -- (-0.95,0.23) -- (-1.4,0.7)--(-1.2,1.4) --cycle;

%E field
\begin{scope}[rounded corners=12pt, thick,triangle 90 reversed-triangle 90,blue]

\draw (-6,4)-- (-4.5,4) -- (-3,3.5) -- (3,3.5) -- (4.5,4)-- (6,4) ;
\draw (-6,3) -- (-5,3) -- (-4,2.4) ;
\draw (3,2.4) -- (4,3) -- (6,3) ;
\draw (1.5,-0.5) -- (4,-1.5) -- (6,-1.5) ;

\draw (-6,-2) -- (-4,-2) -- (-2.1,-1.2);
\draw (-6,-3) -- (-4.5,-3) -- (-3,-2.5)-- (3,-2.5) -- (4.5,-3)-- (6,-3) ;

\node at (3,4) {$\vec{E}$};
\node at (3,-2.) {$\vec{E}$};
\end{scope}

%mesh1
%\draw  (-1,0) -- (-0.8,0.4) -- (-1.5,0.4)--cycle;

\draw [green,fill=green] (-1.5,0.4) -- (-1.2,-1) -- (-0.7,-0.4)--cycle;
\draw [red,-triangle 90, ultra thick] (0,0) -- (-1.1,-.5 );
\node at (-1.4,-0.65) {\Huge{$\vec{x}$}};

\draw [yellow,fill=yellow] (2,0.8) -- (1.8,2) -- (0.5,0.8)--cycle;
\draw [red,-triangle 90, ultra thick] (0,0) -- (1.2,1.1);

\node at (1.65,1.45) {\Huge{$\vec{x}'$}};
\end{tikzpicture}
\caption{A wire antenna with arbitrary shape located in an applied electric field $\vec{E}$ oriented in the $\mathbf{x_2}$ direction. Position of source and observation mesh are denoted by $\vec{x}'$ and $\vec{x}$.} 
\end{figure}

After finding $\rho_{j}$ over $S$, the polarizability $\gamma_{ij}$ in the $\hat{x}_{i}$ direction due to the applied field in the $\hat{x}_{j}$ direction is 
\begin{align}
\gamma_{ij} = \iint\limits_{S}^{ } x_i \rho_j \left( x \right) dS.
\end{align}
The polarizability $\gamma_{ij}$ has the units $\mathrm{m^{2}V^{-1}}$.  However, it is common to use the normalised polarizability $\gamma_{ij}/a^{3}$ which has units $\mathrm{m^{-1}V^{-1}}$ when $a$ is defined as the radius of the smallest surrounding sphere.

The aim of the \textsc{MoM} method is to convert \eqref{eq_laplace} to the standard matrix form $L\rho_{j}=g$ using a discrete mesh and the summation of basis functions.
Because of their simplicity, pulse functions $f_{n}$ are used for both basis and testing functions: 
\begin{align}
f_{n}=\left\{\begin{matrix}
1 & \text{on} \:  \Delta S_{n}\\ 
0 & \text{otherwise}
\end{matrix}\right.
\end{align}

This choice of basis and testing functions yields the following double integral over the mesh elements for the matrix elements $L_{mn}$ \cite{Arcioni_1997_MTT} 
\begin{align}
L_{mn}= \frac{1}{4 \pi} \iint_{A_m}^{ } dS^{'} \iint_{A_n}^{ } dS \frac{1}{\left | {\vec{x}} -\vec{x}' \right |} ,
\label{eq_MoM-matrix}
\end{align}
where $A_{m}$ and $A_{n}$ are the areas of the $m$th and $n$th triangular mesh (source and observation mesh), respectively. 
For non-diagonal elements, one can write $L_{mn}$ in the approximate form
\begin{align}
L_{mn} \approx \frac{1}{4\pi} \frac{A_n A_m}{\left | {\vec{x}} - \vec{x}' \right |}.
\label{eq_l_mn}
\end{align}

The matrix $L$ shows singular behaviour on the diagonal elements.
This is because the denominator in \eqref{eq_laplace} goes to zero when $\vec{x}$ approaches $\vec{x}'$. 
For diagonal elements, an exact solution to the integral in \eqref{eq_MoM-matrix} can be obtained
from equation (25) in \cite{Eibert_1995_TAP}: 
\begin{align}
L_{nn}&= \frac{A_n^2}{4\pi} \nonumber \\
 & \Bigg\{ 
\frac{1}{6\sqrt{a}}
\log\left[ \frac{\left(a-b+\sqrt{a} \sqrt{a-2b+c} \right) \left( b + \sqrt{a} \sqrt{c} \right)}{\left( -b+\sqrt{a} \sqrt{c} \right) \left( -a+b+\sqrt{a} \sqrt{a-2b+c} \right) } \right]  \nonumber \\
 & +  \left.\frac{1}{6\sqrt{c}}
\log\left[ \frac{ \left( b + \sqrt{a} \sqrt{c} \right)\left(-b+c+\sqrt{c} \sqrt{a-2b+c} \right)}{\left( -b+\sqrt{a} \sqrt{c} \right) \left( b-c+\sqrt{c} \sqrt{a-2b+c} \right) } \right] \right. \nonumber  \\
 & +  \frac{1}{6\sqrt{a-2b+c}}
\log\left[ \frac{\left(a-b+\sqrt{a} \sqrt{a-2b+c} \right) \left( -b +c+ \sqrt{c} \sqrt{a-2b+c} \right)}{\left( b-c+\sqrt{c} \sqrt{a-2b+c} \right) \left( -a+b+\sqrt{a} \sqrt{a-2b+c} \right) } \right]
 \Bigg\}, \label{eq_l_nn}
\end{align}
where $a$, $b$, and $c$ are computed from position vectors of mesh vertices $\vec{r}_{1}$, $\vec{r}_{2}$, and $\vec{r}_{3}$ by 
\begin{align}
a = \left(\vec{r}_3-\vec{r}_1 \right)\cdot\left(\vec{r}_3-\vec{r}_1 \right),\\
b = \left(\vec{r}_3-\vec{r}_1 \right)\cdot\left(\vec{r}_3-\vec{r}_2 \right),\\
c = \left(\vec{r}_3-\vec{r}_2 \right)\cdot\left(\vec{r}_3-\vec{r}_2 \right).
\end{align}

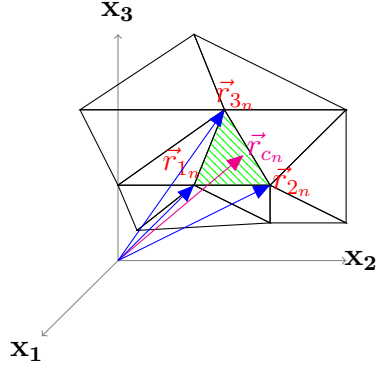
\begin{figure}
\centering

\begin{tikzpicture}
%coordinate system
\draw [help lines,->] (0,0) -- (-1,-1);
\node at (-1.2,-1.2) {$\mathbf{x_1}$};
\draw [help lines,->] (0,0) -- (3,0);
\node at (3.2,0) {$\mathbf{x_2}$};
\draw [help lines,->] (0,0) -- (0,3);
\node at (0,3.3) {$\mathbf{x_3}$};

%mesh1
\draw (1,1) -- (0,1) -- (1.4,2)-- cycle;
\draw (1,1) -- (0,1) -- (0.25,0.4)-- cycle;
\draw (-0.5,2) -- (0,1) -- (1.4,2)-- cycle;
\draw (-0.5,2) -- (1,3) -- (1.4,2)-- cycle;
\draw (3,2) -- (1,3) -- (1.4,2)-- cycle;
\draw (2,1) -- (3,2) -- (1.4,2)-- cycle;
\draw (2,1) -- (3,2) -- (3,0.5)-- cycle;
\draw (1,1) -- (2,0.5) -- (2,1)-- cycle;
\draw (0.25,0.4) -- (2,0.5) -- (1,1)-- cycle;
\draw (3,0.5) -- (2,0.5) -- (2,1)-- cycle;

\draw [pattern=north west lines,pattern color= green](1,1) -- (2,1) -- (1.4,2)-- cycle;
\draw [magenta,arrows={{}-triangle 45}] (0,0) -- (1.65,1.4);
\node [magenta] at (1.95,1.5) {$\vec{r}_{c_n}$};

\draw [blue,arrows={{}-triangle 45}] (0,0) -- (1,1);
\node [red] at (0.85,1.3) {$\vec{r}_{1_n}$};
\draw [blue,arrows={{}-triangle 45}] (0,0) -- (2,1);
\node [red] at (2.3,1.05) {$\vec{r}_{2_n}$};
\draw [blue,arrows={{}-triangle 45}] (0,0) -- (1.4,2);
\node [red] at (1.55,2.2) {$\vec{r}_{3_n}$};

%\draw [->] (0,0) -- (-0.25,0.8);
\end{tikzpicture}
\caption{Discretization of the 3D conductor into mesh elements. The calculation
of $l_{nn}$ needs $\vec{r}_{1_{n}}$, $\vec{r}_{2_{n}}$, and $\vec{r}_{3_{n}}$,
the position of the vertices of $n$th mesh element, and
$\vec{r}_{c_{n}}$ points to the centre of of the mesh.}
\label{Fig_self_mesh_contribution}
\end{figure}

The $m$th element $g$ is 
\begin{align}
g_m=( x_{j_m}+C_j ) A_m,
\label{eq_g_n}
\end{align}
where $x_{j_m}$ is the projection of the centre of the $m$th mesh along $\mathbf{x_j}$ axis.

As long as the centre of the geometry is located at the coordinate origin, $C_{j}$ is zero. 
Problems arise when object is shifted from the origin and large errors can result, particularly for complicated shapes. 
As far as authors know, there is no published literature which describes how to calculate $C_{j}$ in general.
To find $C_{j}$, we define a complementary parameter $u_m = x_{j_{m}}A_{m}$ and rewrite $g$ as: 
\begin{align}
g = u + C_j A.
\end{align}

The induced charge $\rho_{j}$ on the object is calculated from $L^{-1}g$, or,
\begin{align}
\rho_j = L^{-1} u + C_j L^{-1} A.
\label{eq_induced_charge}
\end{align}
In the \eqref{eq_induced_charge}, $\rho_{j}$, $u$ and $A$ are vectors of numbers while $L$ is the \textsc{MoM} matrix. 
By the charge conservation law, the sum of the induced charge $\rho_{j}$ has to vanish.
Therefore $C_{j}$ is found as: 
\begin{align}
C_j = - \frac{\sum L^{-1} u}{\sum L^{-1} A}.
\label{eq_C_j}
\end{align}

Finally the polarizability is computed from: 
\begin{align}
\gamma_{ij} = (x_{i_1}\dots x_{i_m}\dots x_{i_n}) \Big[ L^{-1} g \Big ].
\label{eq_final_polarizability}
\end{align}
where $x_{i_{n}}$ is the position of the centre of the $n$th mesh element in the $\mathbf{x_{i}}$ direction $(\vec{r}_{c_{n}}\cdot\hat{x}_{i}=x_{i_{n}})$.

\section{Implementation}
\label{sec_Implementation}
\begin{figure}
\centering
\begin{tikzpicture}[scale=0.8, every node/.style={scale=0.8}]
\tikzstyle{longblock} = [rectangle, draw=black, fill=blue!20, minimum height=3em, text width=16em] 
\tikzstyle{arrow} = [single arrow, draw]

\node at (0,0) {Triangular mesh file};
\node (box1) at (0,-1) [rectangle,draw=black,fill=blue!20!white] {Extract vertices, faces, normals};
\node (box2) at (0,-2.5) [longblock] {Calculate diagonal and non-diagonal matrix elements \eqref{eq_l_mn},\eqref{eq_l_nn}};
\node (box3) at (0,-3.75) [rectangle,draw=black,fill=blue!20!white] {Matrix inversion};
\node (box5) at (0,-4.75) [rectangle,draw=black,fill=blue!20!white] {$i,j=1$};
\node (box6) at (0,-6.5) [diamond,draw=black,fill=blue!20!white] {$i \leq 3$};
\node (box7) at (2.2,-6.5) [diamond,draw=black,fill=blue!20!white] {$j\leq 3$};
\node (box8) at (6.,-6.5) [rectangle,draw=black,fill=blue!20!white] {$C_j$ calculation \eqref{eq_C_j}};
\node (box9) at (6,-7.5) [rectangle,draw=black,fill=blue!20!white] {Excitation calculation \eqref{eq_g_n}};
\node (box10) at (6,-8.5) [rectangle,draw=black,fill=blue!20!white] {Polarizability $\gamma_{ij}$ \eqref{eq_final_polarizability}};
\node (box11) at (0,-9.25) [ellipse,draw=black,fill=blue!20!white] {Polarizability Tensor};

\draw [->,thick] (0,-0.25) -- (box1.north);
\draw [->,thick] (box1.south) -- (box2.north);
\draw [->,thick] (box2.south) -- (box3.north);
\draw [->,thick] (box3.south) -- (box5.north);
\draw [->,thick] (box5.south) -- (box6.north);
\draw [->,thick] (box6.east) -- (box7.west);
\draw [->,thick] (box7.east) -- (box8.west);
\draw [->,thick] (box8.south) -- (box9.north);
\draw [->,thick] (box9.south) -- (box10.north);
\draw [->,thick] (box10.west) -- ++(-1.65,0) --(box7.south);

\draw [->,thick] (box7.north) -- ++(0,0.3) -- (0,-5.25);
\draw [->,thick] (box6.south) -- (box11.north);

\node at (-0.2,-7.4) {n};
\node at (1.9,-5.6) {n};

\node at (3.1,-6.2) {y};
\node at (0.8,-6.1) {y};
\end{tikzpicture}
\caption{Program flowchart for the polarization calculation of an arbitrary
conductor shape using an array of triangular mesh elements.}

\label{Fig_Flowchart}
\end{figure}
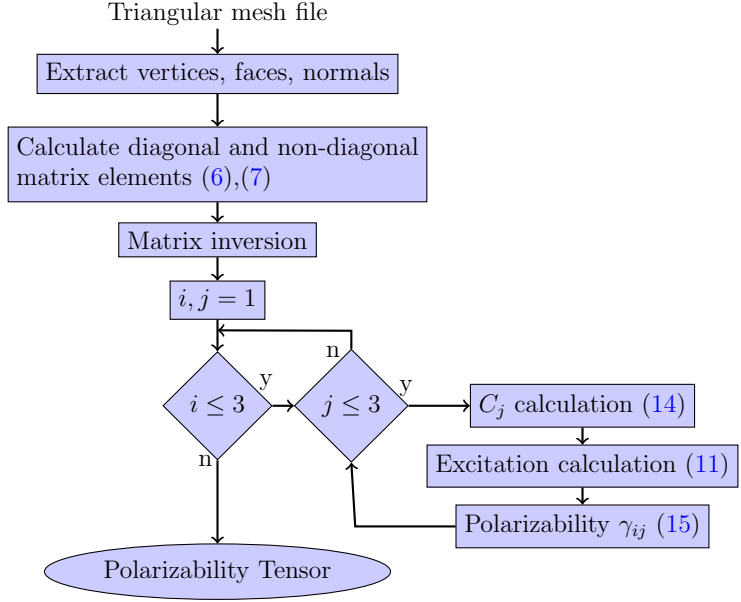

The complete solution was implemented in \textsc{Matlab} code by substituting equations \eqref{eq_l_mn}-\eqref{eq_C_j}
in \eqref{eq_final_polarizability} (see Figure~\ref{Fig_Flowchart}). 
The calculated 3D polarizability is normalised by $a^{3}$, where $a$ is the smallest radius of the sphere that encloses the object.
The mesh file geometry is imported into the \textsc{Matlab} program. 
We used the \textsc{MoM} simulation package \cite{FEKO} to generate the triangular mesh on the 3D objects.
Many other commercial and non-commercial packages can also be used for this mesh generation (e.g., \textsc{AutoCAD} and \textsc{Ansys}).

\section{Validation}
\label{sec_Validation}
The code was validated by comparing the computed results with several published closed form values for simple 3D geometric shapes including a sphere, circular disk,
and toroidal ring~\cite{Gustafsson2009_TAP}. 
Results are normalised by $a^{3}$  and shown in Table~\ref{Table_Compare_Pol}.

\begin{table}
\centering
\caption{Table 1: Comparison of the numerical results with~\cite{Gustafsson2009_TAP}}
\begin{tabular}{|c|c|c|c|}
\hline
Geometry &	Sphere &	Disk &	Toroid\\
\hline
Analytic Result & 12.56 & 5.27 & 2.64\\
\hline
Numerical Result & 12.59 & 5.22 & 2.63\\
\hline
\end{tabular}
\label{Table_Compare_Pol}
\end{table}

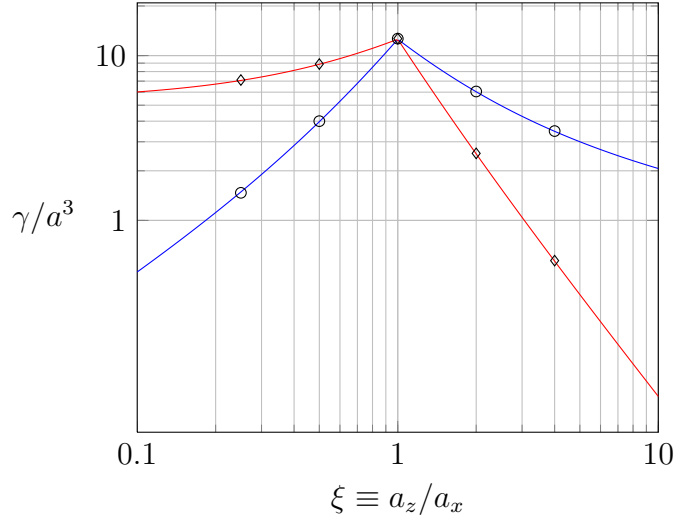
\begin{figure}
\centering
\begin{tikzpicture}
\begin{loglogaxis} [scale=1,xlabel=$\xi \equiv {a_z}/{a_x}$,grid=both, y label style = {rotate=-90},
ylabel=${\gamma}/{a^3}$,xmin = 0.1, xmax=10,
ytick ={1,10},
yticklabel style={/pgf/number format/fixed},
% changes tick labels to a number instead
% of exponential notation:
yticklabel={%
\pgfmathfloatparsenumber{\tick}%
\pgfmathfloatexp{\pgfmathresult}%
\pgfmathprintnumber{\pgfmathresult}%
},
xticklabel style={/pgf/number format/fixed},
% changes tick labels to a number instead
% of exponential notation:
xticklabel={%
\pgfmathfloatparsenumber{\tick}%
\pgfmathfloatexp{\pgfmathresult}%
\pgfmathprintnumber{\pgfmathresult}%
},
]
\addplot [no marks,blue] table [x=xi,y=gammaSV] {gammaSpheroid.txt};
\addplot [no marks,red] table [x=xi,y=gammaSH] {gammaSpheroid.txt};

\addplot [mark=diamond, only marks]coordinates {(0.25,7.1078) (0.5,8.8852) (1,12.6719) (2,2.551) (4,0.5701)};

\addplot [mark=o,only marks] coordinates { (0.25,1.4705) (0.5,4.0017) (1,12.6719) (2,6.0565) (4,3.4847) };
%\legend{$\gamma_V$,$\gamma_H$};
\end{loglogaxis}
%\draw (6,4) rectangle  (6.5,5);
\end{tikzpicture}

\caption{Analytic tangential (red line) and perpendicular (blue line) polarizability
of spheroids with different aspect ratios $\xi ={a_{z}}/{a_{x}}$.
Diamonds and circles refer to our numerical results. When $a_{x}=a_{z}$
the object is a sphere. }

\label{Fig_Pol_Spheroid}
\end{figure}

Figure~\ref{Fig_Pol_Spheroid} shows the polarizability of spheroids of semi-axes $a_x$, $a_y$, $a_z$ in terms of different aspect ratios.
For the oblate and prolate geometries with circular cross sections in $xy$-plane, the red and blue lines illustrate the analytical behaviour of the tangential and perpendicular polarizabilities,
respectively~\cite{Kleinman_Senior_Chapter}. 
Numerical results were computed for three different aspect ratios (1, 2 and 4). 
There is strong agreement with analytical expressions.
A sphere ($a_{x}=a_{z}$) has the highest polarizability and $\gamma_{h}=\gamma_{v}$ at this point. 
A spheroid with an axial ratio of two has a different shape to a spheroid with an axial ratio of 0.5, and so Figure~\ref{Fig_Pol_Spheroid} is not symmetric about the vertical line $a_{x}=a_{z}$.

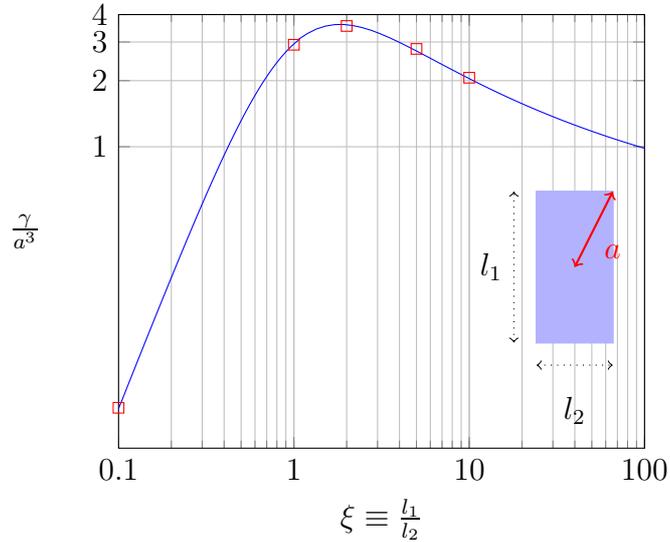
\begin{figure}
\centering
\begin{tikzpicture}
\begin{loglogaxis} [width=8.5cm,scale=1,xlabel=$\xi \equiv \frac{l_1}{l_2}$,grid=both, y label style = {rotate=-90},
ylabel=$\frac{\gamma}{a^3}$,xmin = 0.1, xmax=100, ymax=4,
ytick ={1,2,3,4},
yticklabel style={/pgf/number format/fixed},
% changes tick labels to a number instead
% of exponential notation:
yticklabel={%
\pgfmathfloatparsenumber{\tick}%
\pgfmathfloatexp{\pgfmathresult}%
\pgfmathprintnumber{\pgfmathresult}%
},
xticklabel style={/pgf/number format/fixed},
% changes tick labels to a number instead
% of exponential notation:
xticklabel={%
\pgfmathfloatparsenumber{\tick}%
\pgfmathfloatexp{\pgfmathresult}%
\pgfmathprintnumber{\pgfmathresult}%
},
]
\addplot [red,only marks, mark =square] coordinates { (0.1,0.0647) (1,2.9149)  (2,3.55) (5,2.7902) (10,2.0621)};

\addplot [blue,domain= 0.1:1.08, samples=40] {\x ^2 * (6.275 + 7.328* \x  - 1.651 * \x ^2) / (1+0.8* \x  + 1.025 * \x  ^2 + 1.242 * \x  ^3) };

\addplot [blue,domain= 0.94:100, samples=40] {4 *pi /3 * ((1-\x^-2)^.5) ^3 / ( ln(1+(1-\x^-2)^.5) + ln(\x) - (1-\x^-2)^.5)   * (1.001 + 18.098 * \x^-1 - 11.42 * \x^-2 + 2.266 *  \x^-3) / (1+ 17.074 * \x^-1 -0.309 *\x^-2 + 24.78 * \x^-3)  };

\end{loglogaxis}
%\draw (6,4) rectangle  (6.5,5);
\begin{scope} [scale = 2, xshift = -3.25cm, yshift = -2.3cm ]
\draw [fill,blue!30!white,thick](6,3) rectangle (6.5,4);
\draw [<->,dotted] (6,2.85) -- (6.5,2.85);

\node at (6.25,2.55) {$l_2$};

\draw [<->,dotted] (5.85,3) -- (5.85,4);
\node at (5.7,3.5) {$l_1$};

\draw [<->,thick,red] (6.25,3.5) -- (6.5,4);
\node [red,thick] at (6.5,3.6) {$a$};
\end{scope}
%\draw [dashed] (6.25,3.5) circle [radius=0.65];
\end{tikzpicture}
\caption{Normalised vertical polarizability of infinitely thin rectangles with different length to width ratios (continuous line). 
The rectangles indicate the aspect ratio corresponding to the calculated values (red squares).
An aspect ratio of unity is square.}
\label{Fig_Pol_Rect}
\end{figure}

Figure~\ref{Fig_Pol_Rect} shows the polarizability of several infinitely thin (i.e., 2D), perfectly conducting rectangles with different aspect ratios. 
These results are compared those of Gustafsson~\cite{Gustafsson_2011_LAPC}.
When the aspect ratio is unity the object is square. 
The two sets of results show good agreement.

\section{Discussion and conclusion}
\label{sec_Conclusion}
A numerical technique to compute the polarizability of conducting bodies with arbitrary shapes was reported. 
A desktop computer with Intel\textregistered{}core i5 CPU and 4GB of RAM was used for the numerical calculations. 
The program is fast (less than a minute) for geometries with less than 3000
mesh elements. 
However, the calculation time increases with the cube of mesh elements. 
For example, the computational time is 4 and 14 minutes for geometries with 6000 and 7000 mesh elements, respectively. 
In addition to the code, a GUI interface is available from the authors.

The utilisation of other basis and test functions (i.e: \textsc{RWG} basis functions) is recommended to improve this technique and obtain more accurate results. 
This investigation used the library \textsc{Matlab} functions, but faster and more efficient methods maybe available for the matrix inversion.

\paragraph{Acknowledgements}
Morteza Shahpari appreciates the Griffith Postgraduate Research Scholarship from Griffith University. 
This work is partly funded by a grant from Australian Research Council DP130102098.

% Preferably provide your bibliography as a separate .bbl file.
% Include DOIs (preferred), URLs, Math Review numbers or Zentralblatt numbers 
% in your bibliography so we help connect the web of science 
% and ensure maximum visibility for your article.

% \printbibliography[title={References}]
\bibliographystyle{plain}
%\bibliography{CTAC2012_BIB} 

\end{document}